\documentclass[a4paper,conference]{IEEEtran}
\usepackage[latin1]{inputenc}
\usepackage[english]{babel}
\usepackage{dsfont}
\usepackage{amsfonts}
\usepackage{amssymb}
\usepackage{amsmath}
\usepackage{mathrsfs}
\usepackage{xcolor}
\usepackage{graphicx}
\usepackage{mathdots}
\usepackage{float}
\usepackage{stmaryrd}
\usepackage[colorinlistoftodos]{todonotes}
\usepackage[ruled,vlined]{algorithm2e}
\usepackage{verbatim}
\usepackage{color}
\usepackage{lipsum}
\usepackage{mathtools}
\usepackage{paralist}   

 \usepackage{url}

\pdfoutput=1
\newtheorem{theo}{Theorem}
\newtheorem{prop}{Proposition}
\newtheorem{lem}{Lemma}

\newtheorem{rem}{Remark}
\newtheorem{defi}{Definition}

\makeatletter
\renewcommand*\env@matrix[1][*\c@MaxMatrixCols c]{%
  \hskip -\arraycolsep
  \let\@ifnextchar\new@ifnextchar
  \array{#1}}
\makeatother

\IEEEoverridecommandlockouts

\title{Fixed-Length Strong Coordination}

\author{ 
  \IEEEauthorblockN{Giulia~Cervia, Tobias~Oechtering, and Mikael~Skoglund}
  
  \IEEEauthorblockA{School of Electrical Engineering and Computer Science,
                    KTH Royal Institute of Technology,\\
                    Stockholm, Sweden,
                    \{cervia, oech, skoglund\}@kth.se}
  \thanks{This work was supported in part by the Swedish foundation for strategic research and the Swedish research council.}
}

\begin{document}
\maketitle

 \begin{abstract} 
We consider the problem of synthesizing joint distributions of signals and actions over noisy channels in the finite-length regime.
For a fixed blocklength $n$ and an upper bound on the distance $\varepsilon$, a coding scheme is proposed such that the induced joint distribution is $\varepsilon$-close in $L^1$ distance to a target i.i.d. distribution. 
The set of achievable target distributions and rate for asymptotic strong coordination can be recovered from the main result of this paper by having $n$ that tends to infinity.
 \end{abstract}


\section{Introduction}

The problem of cooperation of autonomous devices in a decentralized  network, initially raised in the context of game theory by  \cite{gossner2006optimal}, has been introduced in information theory in \cite{cuff2009thesis}. Instead of using the channel between the agents to convey information reliably, coordination is intended as a way to induce a prescribed behavior. Two metrics to measure the level of coordination have been defined: \emph{empirical coordination}, which requires the empirical distribution
of the actions to approach a target distribution with high probability, and \emph{strong coordination}, which requires the $L^1$ distance of the distribution of sequences of actions to converge to an i.i.d. target distribution~\cite{cuff2009thesis}.

While a number of works have studied  the strong coordination region with error free links, namely~\cite{cuff2010}, only a few works have focused on coordination with noisy channels.  However, since in a realistic scenario the communication links are usually noisy, and the signals exchanged over the physical channel are a part of what can be observed, we investigate joint strong coordination of signals and actions over a noisy links as in~\cite{Cervia2017,cervia2018journal}. 

We consider a two-node network composed of an information source and a noisy channel, in which both nodes have access to a common source of randomness. Although the exact coordination region is still unknown, \cite{Cervia2017} presents an inner and an outer bound for the region.
For the inner bound derived in \cite{Cervia2017}, \cite{cervia2018journal} proposes a practical polar coding scheme which achieves strong coordination. However, the computational cost of polar codes makes them impractical for delayed-constraint applications because it involves a chaining construction over a large number of blocks. 

For this reason, we investigate strong coordination in the finite-length regime for the same point-to-point setting of \cite{Cervia2017, cervia2018journal}, introducing the notion of \emph{fixed-length strong coordination}.
Using the finite-length techniques of \cite{polyanskiy2010channel,kostina2012fixed}, combined with the random binning approach inspired by \cite{yassaee2013non}, we present an inner bound for the fixed-length strong coordination region.
We develop a joint source-channel scheme in which an auxiliary codebook allows us to jointly coordinate signals and actions.

The rest of the document is organized as follows. 
$\mbox{Section \ref{sec: sys}}$ introduces the notation and some preliminary results, and describes the model under investigation.  In particular, the results on strong coordination in the asymptotic setting are recalled.
Then, $\mbox{Section \ref{sec: nonasy}}$ studies the problem of strong coordination in the non-asymptotic setting, and derives an inner bound for the fixed-length strong coordination region.


\section{System model and background}\label{sec: sys}

\subsection{Preliminaries}\label{subsec: prel}
We define the integer interval $\llbracket a,b \rrbracket$ as the set of integers between $a$ and $b.$
We use the notation ${\lVert \cdot \rVert}_{1}$  and $\mathbb D (\cdot \Arrowvert \cdot)$ to denote the $L^1$  distance  and Kullback-Leibler (K-L) divergence respectively.  

We recall some useful definition and results.

\begin{defi}
Given $A \sim P_A$ and $(A,B) \sim P_{AB}$
\begin{itemize}
\item Information: $ h_{P_A}(a)\coloneqq \log{\frac{1}{P_{A}(a)}}$;
\item Conditional information:
$h_{P_{A|B}} (a|b)\coloneqq \log{\frac{1}{P_{A|B}(a|b)}} $;
\item Information density: $\imath_{P_{AB}}(a,b)\coloneqq \log{\frac{P_{AB}(a,b)}{P_{A}(a) P_{B}(b)}}$.
\end{itemize}
\end{defi}

\vspace{1mm}

\begin{lem}[Properties of $L^1$  distance]\label{tv prop}
\begin{enumerate}[(i)]
\item \label{cuff16}${\lVert P_{A}\!- \!\hat P_{A}\rVert}_{1} \!\leq\! {\lVert P_{AB}\!-\! \hat P_{AB}\rVert}_{1}$, see \cite[Lemma 16]{cuff2009thesis},
\item  \label{cuff17}${\lVert P_A\!- \!\hat P_A\rVert}_{1}\!\!\!=\!\! {\lVert P_AP_{B|A}\!-\! \hat P_A P_{B|A}\rVert}_{1}$, see \cite[Lemma 17]{cuff2009thesis},
\item \label{lem4} If ${\lVert P_{A} P_{B|A}\!-\! P'_{A} P'_{B|A}\rVert}_{1}$ $ = \varepsilon$,  then 
there exists $a \in \mathcal A$ such that 
${\lVert P_{B|A=a}\!-\!P'_{B |A= a}\rVert}_{1}\! \leq\! 2 \varepsilon$, see \cite[Lemma 4]{yassaee2014achievability}.

\end{enumerate}
\end{lem}

\begin{defi}\label{defcoup}
A coupling of  $P_A$ and $P_{A'}$ on  $\mathcal A$ is any  
$\hat P_{AA'}$ on $\mathcal{A} \times \mathcal{A}$  whose marginals are $P_A$ and $P_{A'}$.
\end{defi}

\begin{prop}[Coupling property $\mbox{\cite[I.2.6]{Lindvall1992coupling}}$]\label{theocoup}
Given $A \sim P_{A}$, $A' \sim P_{A'}$, any coupling 
$\hat P_{AA'}$ of $P_{A}$, $P_{A'}$ satisfies
\begin{equation*}
{\lVert P_A- P_{A'}\rVert}_{1}\leq 4 \, \mathbb P_{\hat P_{AA'}}\{A \neq A'\}.
\end{equation*}
\end{prop}


\begin{center}
\begin{figure}[t]
\centering
\includegraphics[scale=0.17]{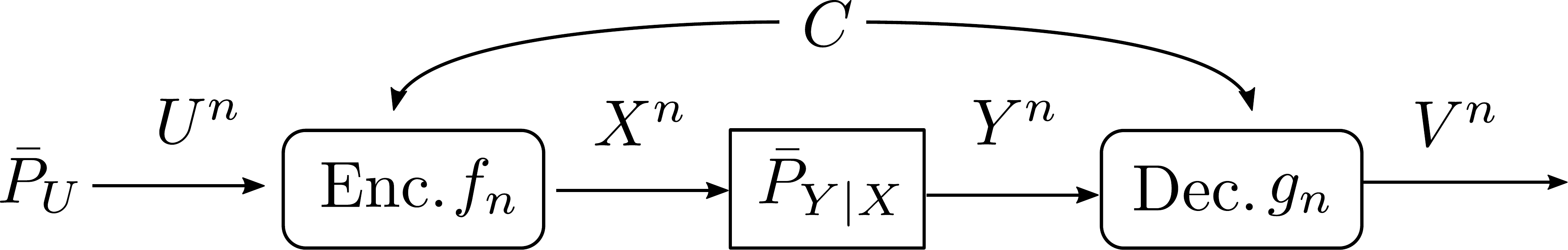}
\vspace{-1mm}
\caption{Point-to-point joint strong coordination setting.}
\label{fig: isit2017}
\vspace{-5mm}
\end{figure}
\end{center}

\vspace{-8mm}
\subsection{Point-to-point setting}\label{subsecsec: sys}
We consider the two-nodes network of Figure \ref{fig: isit2017}, comprised of an i.i.d. source with distribution $\bar P_{U}$, and a discrete memoryless channel  $\bar P_{Y |X}$.
Two agents, an encoder and a decoder, share a source of common randomness $C \in \llbracket 1,2^{nR_0} \rrbracket$.
The encoder selects a signal $X^{n}= f_n(U^{n},  C)$,  $f_n: \mathcal U^n  \times  \llbracket 1,2^{nR_0} \rrbracket \rightarrow \mathcal X^n$, which is is transmitted over $\bar P_{Y |X}$.
The decoder observes  $Y^{n}$ and common randomness $C$, and it selects an action $V^{n} = g_n(Y^{n}, C)$, $g_n: \mathcal Y^n  \times \llbracket 1,2^{nR_0} \rrbracket \rightarrow \mathcal V^n$.
For block length $n$, the pair $(f_n , g_n )$  constitutes a code.

\vspace{1mm}
\subsection{Asymptotic case}
\vspace{1mm}
In the asymptotic regime, a pair $(\bar{P}_{UXYV}, R_0)$ is achievable for strong coordination  for the setting of Figure \ref{fig: isit2017} if  
\vspace{0.5mm}
\begin{equation*}
\lim_{n \to \infty} {\lVert P_{U^{n}  X^{n} Y^{n}  V^{n}}- \bar{P}_{UXYV}^{\otimes n} \rVert}_{1}=0,
\end{equation*}where $P_{U^{n} X^{n} Y^{n}  V^{n}}$ is the joint distribution induced by the code, and the strong coordination region is the closure of the set of achievable  $(\bar{P}_{UXYV}, R_0)$ (see \cite{cuff2010}).

While the joint strong coordination region of signals ($X^n$ and $Y^n$) and actions ($U^n$ and $V^n$) is still unknown, the best known inner bound is derived in \cite[Thm.~1]{Cervia2017}:
\vspace{1mm}
\begin{equation}\label{eq: regionisit}
\begin{Bmatrix}[l]
(\bar P_{UXYV},R_0) : \\
\quad \bar P_{UXYV}=\bar P_{U}  \bar P_{X|U} \bar P_{Y|X} \bar P_{V|UXY} \\
\quad \exists \, W\in\mathcal{W}, \, W \sim\bar P_{W|UXYV} \textnormal{ s.t. }\\ 
\quad  \bar P_{UWXYV}=\bar P_{U} \bar P_{W|U} \bar P_{X|UW} \bar P_{Y|X} \bar P_{V|WY}\\
\quad I(W;U) \leq I(W;Y)\\
\quad R_0 \geq I(W;UXV|Y)
\end{Bmatrix}\!.
\end{equation}

\vspace{1mm}
\section{Non-asymptotic case }\label{sec: nonasy}
\vspace{0.5mm}

We introduce the notion of fixed-length strong coordination.
\begin{defi}[Fixed-length strong coordination]\label{def strong coord fix len}
A pair $(\bar{P}_{UXYV}, R_0)$ is $(\varepsilon,n)$-achievable for strong coordination if for a fixed $n>0$,
there exists $\varepsilon>0$ and a code $(f_n,g_n)$ with common randomness rate $R_0$, such that
$${\lVert P_{U^{n}  X^{n} Y^{n}  V^{n}}- \bar{P}_{UXYV}^{\otimes n}  \rVert}_{1}\leq \varepsilon,$$
where $P_{U^{n} X^{n} Y^{n}  V^{n}}$ is the joint distribution induced by the code.
Then, the  \emph{fixed-length strong coordination region} $\mathcal R$ is the closure of the set of achievable  $(\bar{P}_{UXYV}, R_0)$.
\end{defi}

\vspace{1mm}
For the setting of Figure \ref{fig: isit2017}, the main result of this paper is the following inner bound for the fixed-length $n$ strong coordination region when $\varepsilon$ is a multiple of $1/\sqrt n$. 

\vspace{1mm}

\begin{theo}[Inner bound]\label{theona}
Let $\bar P_{U}$ and $\bar P_{Y|X}$ be the given source and channel distributions, then 
$\mathcal R_{\text{in}} \!\subseteq $ $\mathcal R$:
\vspace{0.5mm}
\begin{equation}\label{region fn}
\mathcal R_{\text{in}} \! \coloneqq\!\!\begin{Bmatrix}[l]
(\bar P_{UXYV},R_0) :\\[0.5mm]
\quad \bar P_{UXYV}=\bar P_{U} \bar P_{X|U}\bar P_{Y|X} \bar P_{V|UXY}   \\[0.4mm]
\quad \exists \,W\in\mathcal{W},\, W \sim\bar P_{W|UXYV} \textnormal{ s.t. }\\[0.4mm]
\quad \bar P_{UWXYV}=\bar P_{U} \bar P_{W|U} \bar P_{X|UW} \bar P_{Y|X} \bar P_{V|WY}\\[1mm]
\quad I(W;U) \leq I(W;Y) + c_1 \varepsilon  +  O( \frac{\log n}{n}) \\[0.4mm]
\phantom{\quad I(W;U) +} + Q^{-1} \left(c_2 \varepsilon +O( \frac{1}{\sqrt n} ) \right) \sqrt{ \frac{V_{\bar P}}{n}}  \\[1mm]
\quad R_0 \geq  I(W;UXV|Y) +c_3  \varepsilon+  O(\! \frac{\log n}{n}\! ) \\[0.4mm]
\phantom{\quad R_0 +} + Q^{-1} \left(c_2  \varepsilon + O( \frac{1}{\sqrt n})\right) \sqrt{  \frac{V_{\bar P}}{n}}
\end{Bmatrix}
\end{equation}where the constants $(c_1, c_2, c_3)$ are defined in \eqref{final rate}, $Q(t)= \int_{t}^{\infty} \frac{1}{\sqrt{2 \pi}} e^{-x^2/2} dx$  is the tail distribution function of the standard normal distribution, and $V_{\bar P} $ is the dispersion of the channel $\bar P_{Y|W}$ as defined in \cite[Thm.~49]{polyanskiy2010channel}.\end{theo}


\paragraph*{Outline of the proof of Theorem \ref{theona}}
The achievability proof is based on  non-asymptotic \emph{ output statics of random binning} \cite{yassaee2013non} and requires the following steps:
\begin{itemize}
\item[A.] preliminary definitions and results on random binning;
\item[B.] two schemes are defined  for the one-shot problem, a random binning and a random coding scheme;
\item[C.] the scheme is generalized  for a fixed $n$, by repeating the  one-shot scheme $n$ times. Using the properties of random binning, it is possible to derive an upper bound on the  $L^1$ distance between the i.i.d. random binning distribution $P^{\text{RB}}$ and random coding distribution $P^{\text{RC}}$. With common randomness greater than $H(W|Y)+\text{constant} \cdot \varepsilon$, a first bound on ${\lVert P^{\text{RB}} - P^{\text{RC}} \rVert}_{1} $ is found.
Then, a second bound $\varepsilon_{\text{Tot}}$ is recovered, by reducing the rate of common randomness to obtain the conditions in \eqref{region fn}.
\item[D.] the term $\varepsilon_{\text{Tot}}$ is analyzed;
\item[E.] the rate conditions are summarized.
\end{itemize}

\vspace{0.5mm}
 \begin{rem}
Observe that, as we will see in Section \ref{subsec: reduce rate cr},  the final  bound $\varepsilon_{\text{Tot}}$ on  the $L^1$ distance between $P^{\text{RB}}$ and $P^{\text{RC}}$ is worst than the one found in Section \ref{subsec: before reducing}.
However, by worsening the $L^1$ distance, we reduce the rate of common randomness.
 \end{rem}

\vspace{0.5mm}
\begin{rem}[Comparison with the asymptotic case]
Note that, for both the asymptotic and the fixed-length case, the decomposition of the target joint distribution is the same (see~\eqref{eq: regionisit} and~\eqref{region fn}). Perhaps more interestingly,  in the asymptotic regime $\varepsilon$ vanishes, and 
\begin{equation}\label{lognn}
\text{constant}  \cdot \varepsilon+ O \left(\! \frac{\log n}{n} \! \right)\!+ Q^{-1}\!(\text{constant} \cdot \varepsilon + O( \frac{1}{\sqrt n})) \sqrt{ \frac{ V_{\bar P}}{n}}
\end{equation}goes to zero since $\log n/n $ goes to zero and so does the last term because 
{\allowdisplaybreaks
\begin{align*}
  \sqrt{ V_{\bar P}} \frac{Q^{-1}(\text{constant} \cdot \varepsilon + O( \frac{1}{\sqrt n}))}{\sqrt n}  \sim \sqrt{ V_{\bar P}} \frac{\log{\sqrt{n}}}{\sqrt n}\to 0.
\end{align*}}Hence, we can recover the inner bound for the asymptotic region of \eqref{eq: regionisit} from the inner bound for the fixed-length  \eqref{region fn}. 
Moreover, the bound $\varepsilon_{\text{Tot}}$ on the  $L^1$ distance between the two distribution goes to zero as $1/\sqrt n$, as we will see in \eqref{speed conv}.
\end{rem}


\subsection{Preliminaries on random binning}\label{osrb properties}
Let $(A,B)$ $\sim$ $P_{AB} $ be a discrete source and 
$\varphi:$ $ \mathcal A$ $\to$ $\! \llbracket 1,2^{R} \rrbracket,$ $a \mapsto k$, be a uniform random binning of $A$, with $K\coloneqq \varphi(A)$.
We denote the distribution induced by the binning as
\begin{equation}\label{prb1}
 P^{\text{RB}} (a,b,k )\coloneqq P_{AB}(a,b) \mathds 1 \{\varphi(a)=k \}.
\end{equation}

The first objective consists of ensuring that the binning is almost uniform and almost independent from the source
so that the random binning scheme and the random coding scheme generate joint distributions that have the same statistics.
\vspace{0.5mm}
\begin{theo}[$\mbox{\cite[Thm.~1]{yassaee2013non}}$]\label{oneshot1}
 Given $P_{AB}$, for every distribution $T_B$ on $\mathcal B$ and any $\gamma \in \mathbb R^{+}$, $P^{\text{RB}}$ the marginal of \eqref{prb1}
 satisfies
 {\allowdisplaybreaks
 \begin{align}
  &\mathbb E {\lVert  P^{\text{RB}} (b,k) - Q_{K}(k) P_B(b) \rVert}_{1} \leq \varepsilon_{\text{App}}, \nonumber\\
  & \varepsilon_{\text{App}}\coloneqq P_{AB}{ \left( \mathcal S_{\gamma_1} (P_{AB} \| T_B )^{\mathrm{c}}  \right)}+2^{-\frac{\gamma+1}{2}},\label{eqos1}\
 \end{align}}where for a set $X$, we denote with $Q_X$ the uniform distribution over $\mathcal X$ and
\begin{equation}
 \mathcal S_{\gamma}(P_{AB} \| T_B )\!\coloneqq\! \left\{  (a,b) : h_{P_{AB}}\!(a,b)\!-\! h_{T_B} \!(b) \!-\! nR \!> \!\gamma \right\}\!.
\end{equation}
\end{theo}

Before stating the second property,  we recall the definition of a \emph{mismatch stochastic likelihood coder (SLC)}.
\begin{defi}[Mismatch SLC]
Let $T_{AB}$ be an arbitrary probability mass function, and  $\varphi\!: \! \mathcal A\! \to\! \llbracket 1,2^{R} \rrbracket$, $a \mapsto k$ a uniform random binning of $A$. A  mismatch SLC is defined by the following induced conditional distribution
 \begin{equation}
 \hat T_{\hat A|BK} (\hat a|b,k)\coloneqq \frac{T_{A|B}(\hat a|b) \mathds 1 \{\varphi(\hat a)=k \}}{\sum_{\bar a \in \mathcal A} T_{A|B}(\bar a|b) \mathds 1 \{\varphi(\bar a)=k \}}.
\end{equation}\end{defi}

Then, the following result is used to bound the error probability of decoding $A$ when the decoder has access to the side information $B$ as well as to bin indices $\varphi(A)=K$.
\begin{theo}[$\mbox{\cite[Thm.~2]{yassaee2013non}}$]\label{oneshot2}
 Given $P_{AB}$ and any distribution $T_{AB}$, the following bound on the error probability of mismatch SLC holds
 \begin{equation}\label{eqos2}
  \mathbb E \left[ P[\mathcal E] \right] \leq P_{AB} {\left(\mathcal S_{\gamma_1} (T_{AB})^{\mathrm{c}} \right)} + 2^{-\lvert \gamma \rvert} =: \varepsilon_{\text{Dec}},
 \end{equation}where $\gamma$ is an arbitrary positive number and 
 \begin{equation}
  \mathcal S_{\gamma} (T_{AB})\coloneqq \left\{ (a,b) : nR -h_{T_{A|B}} (a|b) > \gamma \right\}.
 \end{equation}
\end{theo}


\begin{center}
\begin{figure}[t]
\centering
\includegraphics[scale=0.17]{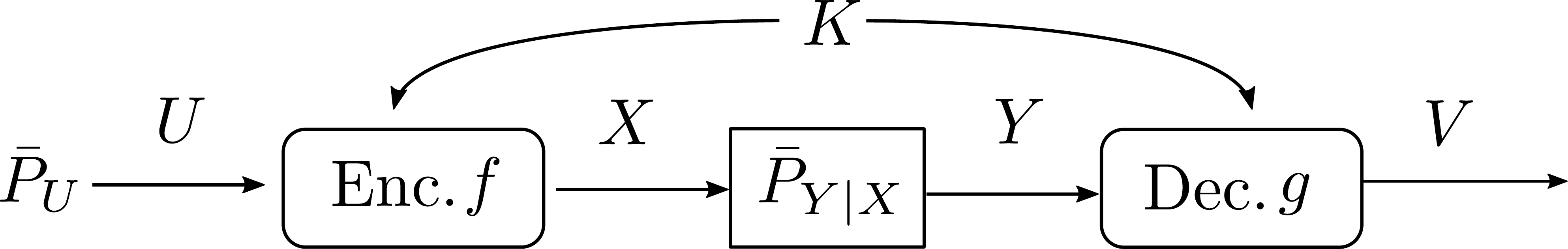}
\vspace{-1mm}
\caption{One-shot system model.}
\label{fig: oneshot}
\vspace{-0.5cm}
\end{figure}
\end{center}

\vspace{-0.8cm}
\subsection{One-shot coordination scheme}\label{one shot scheme}
We consider the setting of Figure \ref{fig: oneshot}.
The encoder and the decoder share a source of uniform randomness $K \in \llbracket 1,2^{R_0} \rrbracket$. The encoder observes the source  $U \in \mathcal U$  and selects a signal  $X= f(U,  K)$,  $f: \mathcal U  \times  \llbracket 1,2^{R_0} \rrbracket \rightarrow \mathcal X$, which is then transmitted over the discrete channel $\bar P_{Y |X}$.
Then, the decoder selects an action $V= g(Y, K)$, where $g: \mathcal Y  \times \llbracket 1,2^{R_0} \rrbracket \rightarrow \mathcal V$.  

\vspace{0.5mm}
\subsubsection{Random binning scheme}

Let $ \bar P_{UXYV}$ be the target distribution,
\begin{equation*}
 \bar P_{U}  \bar P_{X| U} \bar P_{Y|X} \bar P_{V|UXY}.
\end{equation*}
We introduce an auxiliary random variable $W$ such that the sequence $(U$, $X$, $W$, $Y$, $V)$ 
has distribution 
\begin{equation*}
\bar P^{\text{os}}\coloneqq\bar P_{U} \bar P_{W| U} \bar P_{X| W U} \bar P_{Y|X} \bar P_{V|W Y}.
\end{equation*}
We consider two uniform random binnings for $W$:
\begin{enumerate}\setlength{\itemsep}{0.2em}
\item binning $K = \varphi_1(W)$, where  $\varphi_1: \mathcal{W} \to \llbracket 1,2^{R_0} \rrbracket$,  
\item binning $M= \varphi_2(W)$, where $\varphi_2: \mathcal{W} \to \llbracket 1,2^{ R} \rrbracket$,  
\end{enumerate}and the decoder reconstructs $\hat W$ via the mismatch SLC
 \begin{equation*}
  \hat T_{\hat W|YKM} (\hat{w}|y, k, m) \!\!\coloneqq\! \!\frac{\bar P_{W|Y}(\hat{w}|y) \mathds 1 \{\varphi(\hat{w})=(k,m) \}}{\!\sum_{\bar{w} \in \mathcal W} \!\bar P_{W|Y}(\bar{w}|y) \mathds 1 \{\varphi(\bar{w})\!=\! (k, m)\}}.
 \end{equation*}
This induces a joint distribution:
\begin{equation*}
P^{\text{RB,os}}\!\!\coloneqq\!\!\bar P_{U } \bar P_{W|U} \bar P_{X|W U}  \bar P_{K|W} \bar P_{M|W} 
\bar P_{Y|X} \bar P_{V|WY}  \hat T_{\hat W|YKM}.
\end{equation*}In particular, $P^{\text{RB,os}}_{W|KMU}$ is well defined.


\subsubsection{Random coding scheme}
Suppose that in the setting of Figure~\ref{fig: oneshot}, the encoder and decoder have access not only to common randomness $K$ but also to extra randomness $M$, where $K$ is generated uniformly at random  in $\llbracket 1,2^{R_0} \rrbracket$ with distribution $Q_K$ and $M$ is generated uniformly at random in $\llbracket 1,2^{R} \rrbracket$ with distribution $Q_M$ independently of $K$. 
Then, the encoder generates $W$ according to $P^{\text{RB,os}}_{W|KMU}$ defined above, and $X$ according to $\bar P_{X|U W}$.
The encoder sends $X$ through the channel.
The decoder obtains $Y$ and $(K,M)$ and reconstructs $W$ 
via the conditional distribution  $\hat T_{\hat W|YKM}$. 
The decoder then generates $V$ according to the distribution 
$P^{\text{RC,os}}_{V|\hat W Y}(\hat{v}|\hat{w}, y)=\bar P_{V| W Y}(\hat{v}|\hat{ w}, y),$
where $\hat{w}$ is the output of the mismatch SLC.
This defines a joint distribution:
\begin{equation*}
P^{\text{RC,os}}\!\coloneqq\! Q_K Q_M \bar P_U P^{\text{RB,os}}_{W|KMU} \bar P_{X|W U}\bar P_{Y|X} \hat T_{\hat W|YKM} P^{\text{RC,os}}_{V|\hat W Y}.
\end{equation*}

\subsection{Fixed-length coordination scheme}\label{fixed length scheme}
Now, we consider the setting of Figure \ref{fig: isit2017}.
Assume that 
$U^{n}$, $X^{n}$, $W^{n}$, $Y^{n}$ and $V^{n}$ 
are jointly i.i.d. with distribution 
\begin{equation*}
\bar P\coloneqq\left(\bar P^{\text{os}}\right)^{\otimes n}=\bar P_{U^{n}} \bar P_{W^{n}| U^{n}} \bar P_{X^{n}| W^{n} U^{n}} \bar P_{Y^{n}|X^{n}} \bar P_{V^{n}|W^{n} Y^{n}}.
\end{equation*}


\subsubsection{Random binning and random coding scheme}
We repeat the one-shot schemes of Section \ref{one shot scheme} for $n$ i.i.d. uses of the source $\bar P_U^{\otimes n}$ and of the channel $\bar P_{Y|X}^{\otimes n}$:
{\allowdisplaybreaks
\begin{subequations}
\begin{align}
 P^{\text{RB}}\coloneqq &{\left(P^{\text{RB,os}}\right)}^{\otimes n}= P^{\text{RB}}_{U^n  X^n  Y^n  V^n  W^n \hat W^n CF} \nonumber\\
 =& \bar P_{U ^n} \bar P_{W^n|U^n} \bar P_{X^n|W^n U^n}  \bar P_{C|W^n} \bar P_{F|W^n }\nonumber \\ & \bar P_{Y^n|X^n} \hat T_{\hat W^n|Y^n C F} \bar P_{V^n|W^n Y^n }\\[1mm]
 P^{\text{RC}}\coloneqq &{\left(P^{\text{RC,os}}\right)}^{\otimes n}=P^{\text{RC}}_{U^n  X^n  Y^n  V^n  W^n \hat W^n CF}\nonumber \\
 =& Q_{C} Q_{F} \bar P_{U^n} P^{\text{RB}}_{W^n|C F U^n} \bar P_{X^n|W^n U^n}\nonumber\\ & \bar P_{Y^n|X^n} \hat T_{\hat W^n|Y^n C F} P^{\text{RC}}_{V^n|\hat W^n Y^n}
\end{align}\end{subequations}}where $C$ $\coloneqq$ $K^n$, $F\coloneqq M^n$, and for  $T_{W^n|Y^n}\coloneqq \prod_{i=1}^n \bar P_{W|Y} $ the mismatch SLC is:
 {\allowdisplaybreaks
 \begin{align}
  \MoveEqLeft[1.2]
  \hat T_{\hat W^n|Y^n C F} (\hat {\mathbf w}|\mathbf y, \mathbf c, \mathbf f)\nonumber\\[0.7mm]
 &\coloneqq \frac{T_{W^n|Y^n}(\hat {\mathbf w}|\mathbf y) \mathds 1 \{\varphi(\hat {\mathbf w})=( \mathbf c, \mathbf f) \}}{\sum_{\bar {\mathbf w} \in \mathcal W^n} T_{W^n|Y^n}(\bar {\mathbf w}|\mathbf y) \mathds 1 \{\varphi(\bar {\mathbf w})= ( \mathbf c, \mathbf f)\}}\nonumber\\[0.7mm]
 &= \frac{\prod_{i=1}^n \bar P_{W|Y}(\hat w_i|y_i) \mathds 1 \{\varphi(\hat{w_i})=( k_i, m_i) \}}{\sum_{\bar{w_i} \in \mathcal W} \! \prod_{i=1}^n \! \bar P_{W|Y}(\bar{w_i}|y_i) \mathds 1 \{\varphi(\bar{w}_i)\! =\! ( k_i,m_i)\}}.\label{decoder}
 \end{align}}

Observe that the distribution $ P^{\text{RB}}$ is by construction trivially close in $L^1$ distance to the target distribution $\bar P$. 
We use the properties of random binning to show that the random binning and the random coding scheme are $\varepsilon$-close in $L^1$ distance, and therefore so are $ P^{\text{RC}}$ and  $\bar P$.

 \vspace{0.5mm}
\subsubsection{Strong coordination of $(U^n , X^n , Y^n , V^n , W^n )$ --- First bound}\label{subsec: before reducing}
By applying Theorem \ref{oneshot1} and  Theorem \ref{oneshot2} to  $n$ i.i.d. copies of $P^{\text{RB,os}}$ and $P^{\text{RC,os}}$, we have  
{\allowdisplaybreaks
\begin{align*}
  &{\lVert P^{\text{RB}}_{U^n W^n X^n  Y^n   CF}  - P^{\text{RC}}_{U^n W^n X^n  Y^n   CF} \rVert}_{1} \\[0.5mm]
  &\! \overset{\mathclap{(a)}}{=}\! \! {\lVert \bar P_{U^n }\! \bar P_{W^n|U^n}  \! \bar P_{C|W^n} \! \bar P_{F|W^n} \!- \! Q_C Q_F\! \bar P_{U^n} \!P^{\text{RB}}_{W^n|CFU^n}  \! \rVert}_{1}\! \!\! \leq \!\! \varepsilon_{\text{App}},\\[1mm]
  & \mathbb E \left[ P[\mathcal E] \right] \leq \varepsilon_{\text{Dec}}, 
\end{align*}}where $(a)$ comes from (\ref{cuff17}) in Lemma \ref{tv prop}, and
{\allowdisplaybreaks
\begin{subequations}
\begin{align}
 \varepsilon_{\text{App}} &\coloneqq \bar P_{U^n CF}{ \left( \mathcal S_{\gamma_1}^{\mathrm{c}} \right)} +2^{-\frac{\gamma_1+1}{2}},\label{epsilonapp}\\[0.8mm]
 \varepsilon_{\text{Dec}} &\coloneqq \bar P_{W^n Y^n} {\left(\mathcal S_{\gamma_2}^{\mathrm{c}} \right)} + 2^{-\lvert \gamma_2 \rvert}, \label{epsilondec}
\end{align}\end{subequations}}with $\gamma_1$ and $\gamma_2$ arbitrary positive numbers, and 
 {\allowdisplaybreaks
 \begin{subequations}
 \begin{align}
\mathcal S_{\gamma_1}\coloneqq & \, \mathcal S_{\gamma_1} (\bar P_{U^n CF} \|  \bar P_{U^n} )\nonumber \\[0.5mm]
=&\{  ( \mathbf u,\! \mathbf w): h_{\bar P}(\mathbf u, \mathbf w)
 \!-\! h_{ \bar P} (\mathbf u)\! - \!n(R+\!R_0)\!>\! \gamma_1 \! \},\label{sgamma1}\\[1.5mm]
 \mathcal S_{\gamma_2} \coloneqq& \mathcal S_{\gamma_2} (\bar P_{W^n Y^n})\!
=\!\! \{(\mathbf w, \mathbf y) : n(R+R_0)\!-\!h_{\bar P} (\mathbf w|\mathbf y) \!> \gamma_2 \}\nonumber \\
\overset{\mathclap{(b)}}{=}& \Big\{ (\mathbf w, \mathbf y) : n(R+R_0) -\!\sum_{i=1}^n h_{\bar P} (w_i|y_i)\! > \gamma_2\Big\},\label{sgamma2}\vspace{-2mm}
 \end{align}\end{subequations} }where $(b)$ comes from the choice of the mismatch SLC \eqref{decoder}.
Then, we have
{\allowdisplaybreaks
\begin{align*}
& {\lVert P^{\text{RB}}_{U^n W^n X^n  Y^n   CF}  \hat T_{\hat W^n|Y^n C F}\! -\!   P^{\text{RC}}_{U^n W^n X^n  Y^n   CF}  \hat T_{\hat W^n|Y^n C F}   \rVert}_{1} \\[0.5mm]
&= \!{\lVert P^{\text{RB}}_{U^n W^n X^n  Y^n   CF \hat W^n} \!  -\!   P^{\text{RC}}_{U^n W^n X^n  Y^n   CF \hat W^n}   \rVert}_{1}  \leq\!  \varepsilon_{\text{App}}\!+\! \varepsilon_{\text{Dec}}.
\end{align*}}
To conclude, observe that in the random binning scheme we have $V^n \sim   \bar P_{V^n|W^nY^n} $, $W^n \sim \bar P_{W^n|U^n}$, while in the random coding scheme we have $V^n \sim P^{\text{RC}}_{V^n|\hat W^n Y^n}$, $\hat W^n \sim \hat T_{\hat W^n|Y^nCF}$. Then, by Proposition \ref{theocoup},
\begin{equation*}
 {\lVert P^{\text{RB}} - P^{\text{RC}} \rVert}_{1}\leq  \varepsilon_{\text{App}}+5 \, \varepsilon_{\text{Dec}}.
\end{equation*}

\vspace{1mm}
\subsubsection{Reducing the rate of common randomness --- Final bound}\label{subsec: reduce rate cr}
Even though the extra randomness $F$ is required
to coordinate $(U^n , X^n , Y^n , V^n , W^n )$, we do not need it in order to coordinate
only $(U^n , X^n , Y^n , V^n  )$.
We would like to reduce the amount of common randomness by having the two nodes agree on an instance $F = \mathbf f$. To do so, we apply 
Theorem \ref{oneshot1} to $A=W^n$, $B=(U^n, X^n, Y^n, V^n)$, $P_{B}=  P^{\text{RB}}_{U^n X^n Y^n V^n}$, $P_{AB}= P^{\text{RB}}_{U^n X^n Y^n V^n W^n}$ and $K=F$. Then, we have 
 \begin{equation}\label{eq3}
\lVert  P^{\text{RB}}_{U^n X^ nY^n V^n F }   {- Q_{F}P^{\text{RB}}_{U^n X^n Y^n V^n  }  \rVert}_{1} \leq \varepsilon_{\text{App},2},
 \end{equation}where
 \vspace{-1mm}
 {\allowdisplaybreaks
 \begin{subequations}
 \begin{align}
  & \varepsilon_{\text{App},2}\coloneqq
  P^{\text{RB}}{ \left( \mathcal S_{\gamma_3}^{\mathrm{c}} \right)}+2^{-\frac{\gamma_3+1}{2}},\label{epsilonapp2}\\[0.2mm]
 &   \mathcal S_{\gamma_3}\coloneqq\mathcal S_{\gamma_3} (P^{\text{RB}}_{U^n X^n Y^n V^n W^n} \| P^{\text{RB}}_{U^n X^n Y^n V^n} )\label{sgamma3}\\ 
   & =  \mbox{\small $ \{ (\mathbf u,\! \mathbf x,\!\mathbf y, \! \mathbf v, \!\mathbf w)  : h_{P^{\text{RB}}}(\mathbf u,\! \mathbf x,\!\mathbf y,\! \mathbf v,\! \mathbf w) \! -\! h_{P^{\text{RB}}} (\mathbf u,\! \mathbf x,\!\mathbf y,\! \mathbf v)\!-\!nR \!>\! \gamma_3 \}.$} \nonumber
 \end{align}\end{subequations}}
\vspace{-1.5mm}
 
Now, we recall that  by (\ref{cuff16}) in Lemma \ref{tv prop}, we have
{\allowdisplaybreaks
\begin{align}
{\lVert P^{\text{RB}}_{U^n X^n Y^n V^n  F}-P^{\text{RC}}_{U^n  X^n Y^n V^n F} \rVert}_{1} &  \leq  {\lVert P^{\text{RB} }- P^{\text{RC}}  \rVert}_{1} \nonumber \\ & \leq   \varepsilon_{\text{App}}+5\, \varepsilon_{\text{Dec}}.\label{eqtv1}
\end{align}}
\vspace{-1mm}
Combining  \eqref{eq3} and \eqref{eqtv1} with  the triangle inequality, we have
 {\allowdisplaybreaks
 \begin{align*}
 \MoveEqLeft[2]
 { \lVert Q_F P^{\text{RB}}_{U^n X^n Y^n V^n}-Q_F P^{\text{RC}}_{U^n X^n Y^n V^n  } \rVert}_{1}  \\[0.5mm]
  & \leq {\lVert P^{\text{RB}}_{U^n X^n Y^n V^n F }- Q_F P^{\text{RB}}_{U^n X^n Y^n V^n  } \rVert}_{1}  \\[0.5mm] & \quad+ {\lVert P^{\text{RB}}_{U^n X^n Y^n V^n F }- P^{\text{RC}}_{U^n X^n Y^n V^n F } \rVert}_{1}  \\[0.5mm] 
& \leq \varepsilon_{\text{App},2} +\varepsilon_{\text{App}}+5\,\varepsilon_{\text{Dec}}.
 \end{align*}}By (\ref{lem4}) in Lemma \ref{tv prop}, there exists an instance $F= \mathbf f$, such that 
 \vspace{-4.5mm}
 {\allowdisplaybreaks
 \begin{subequations}
 \begin{align}
  & {\lVert P^{\text{RB}}_{U^n X^n Y^n V^n| F=\mathbf f }, P^{\text{RC}}_{U^n X^n Y^n V^n| F=\mathbf f } \rVert}_{1}   \leq \varepsilon_{\text{Tot}},\label{var tot}\\[0.5mm]
  & \varepsilon_{\text{Tot}}\coloneqq2 \,(\varepsilon_{\text{App},2} +  \varepsilon_{\text{App}}+5\, \varepsilon_{\text{Dec}}).\label{epsilon tot}
 \end{align}\end{subequations}}


\subsection{Analysis of the $L^1$ distance}\label{tv analysis}
Substituting \eqref{epsilonapp}, \eqref{epsilondec}, and \eqref{epsilonapp2} into \eqref{epsilon tot}, the bound on the $L^1$ distance in \eqref{var tot} becomes
{\allowdisplaybreaks
\begin{align}
 & \varepsilon_{\text{Tot}}\!=  \! 2 \,\bar P_{U^n CF}{ \left( \mathcal S_{\gamma_1}^{\mathrm{c}}\right)}+ 10\, \bar P_{W^n Y^n} {\left(\mathcal S_{\gamma_2}^{\mathrm{c}}   \right)}+   2\, P^{\text{RB}}{ \left( \mathcal S_{\gamma_3}^{\mathrm{c}}  \right)}\nonumber\\[0.5mm]
 & \qquad + 2 \, \Big(2^{-\frac{\gamma_1+1}{2}} +5 \cdot 2^{-\lvert \gamma_2 \rvert}+ 2^{-\frac{\gamma_3+1}{2}} \Big). \label{epsilon tot}
\end{align}}
We treat separately the terms $\mathcal S_{\gamma_i}^{\mathrm{c}}$, $i=1,2,3$ to 
understand  which rate conditions we have to impose in order to minimize the measure of the sets as a function of  $\gamma_i $, $i=1,2,3$. In a second instance, we choose the parameters $(\gamma_2, \gamma_2, \gamma_3)$ such that $\varepsilon_{\text{Tot}} $  defined above is small.

\vspace{1.6mm}
\subsubsection{Analysis of $\mathcal S_{\gamma_1}^{\mathrm{c}}$} 
To bound $ \bar P_{U^n CF}{ \left( \mathcal S_{\gamma_1}^{\mathrm{c}}\right)}$ we want to find rate conditions such that  $\mathcal S_{\gamma_1} $ includes the typical set $\mathcal T_{\varepsilon_1}^{(n)}$. Observe that if  $ \mathbf u$ is $\varepsilon_1$-typical and $(\mathbf u,\mathbf w)$ are jointly  $\varepsilon_1$-typical,  then
 {\allowdisplaybreaks
 \begin{align*}
  2^{-nH_{\bar P}(U)(1+\varepsilon_1)} \leq \bar P_{U^n }&(\mathbf u) \leq 2^{-nH_{\bar P}(U)(1-\varepsilon_1)}\\
  2^{-nH_{\bar P}(U,W)(1+\varepsilon_1)} \leq \bar P_{U^n W^n}&(\mathbf u, \mathbf w) \leq 2^{-nH_{\bar P}(U,W)(1-\varepsilon_1)}
 \end{align*}which imply
 {\allowdisplaybreaks
\begin{align}
&\! h_{\bar P} (\mathbf u)= - \log{P_{U^n}(\mathbf u)} \leq 
nH_{\bar P}(U)(1+\varepsilon_1),\nonumber\\
&\! h_{\bar P}(\mathbf u, \mathbf w)=- \log {P_{U^n W^n}\!(\mathbf u, \mathbf w)} \geq 
nH_{\bar P}(U,W)(1-\varepsilon_1),\nonumber\\
&\!  h_{\bar P}(\mathbf w|\mathbf u) \!\geq \! n H_{\bar P}(W|U) \!-\! n\,   \underbrace{\varepsilon_1\left(H_{\bar P}(UW)\!+\!H_{\bar P}(U)\right)}_{\varepsilon_2}.\label{epsilon2}
\end{align}}\vspace{-1mm} Suppose we choose $(R,R_0)$ that satisfy
\begin{equation} \label{rateepsilon2}
R+R_0 <  H_{\bar P}(U,W)- H_{\bar P}(U)- \varepsilon_2 -\frac{\gamma_1}{n}.
\end{equation}Then, if for every $\varepsilon_1$-typical sequence, the following chain of inequalities is verified,
\begin{equation}
h_{\bar P}(\mathbf w| \mathbf u) -  \gamma_1 \geq  n H_{\bar P}(W|U) - n \varepsilon_2  -  \gamma_1> n(R+R_0).
\end{equation}Thus, $\mathcal S_{\gamma_1}$ contains the typical set, and there exists a set $\mathcal S$, $\mathcal T_{\varepsilon_1}^{(n)} \subseteq \mathcal S\subseteq \mathcal S_{\gamma_1} $, such that for every $(\mathbf u,\mathbf w) \in \mathcal S$, we have  
\begin{equation}
n H_{\bar P}(W|U)- n \varepsilon_2  >n(R+R_0)+\gamma_1.
\end{equation}Therefore, we have $\mathcal S_{\gamma_1}^{\mathrm{c}}  \subseteq \mathcal S^{\mathrm{c}} $, and since the rate condition \eqref{rateepsilon2} holds, $\mathcal S^{\mathrm{c}}  $ is empty and the measure $ \bar P_{U^n CF}{ \left( \mathcal S_{\gamma_1}^{\mathrm{c}}\right)}$ is zero.

\vspace{2.2mm}
\subsubsection{Analysis of $\mathcal S_{\gamma_3}^{\mathrm{c}}$}
 Similarly to above, $\mathcal S_{\gamma_3}$ contains all the typical sequences, and  $P^{\text{RB}}{ \left( \mathcal S_{\gamma_3}^{\mathrm{c}}  \right)}$ vanishes if
\begin{equation}\label{epsilon3}
R\!< \!H_{\bar P}(W|UXYV) -\underbrace{\varepsilon_1  (H_{\bar P}(UWXYV)+ H_{\bar P}(W))}_{\varepsilon_3}\!- \frac{\gamma_3}{n}.
\end{equation}

\vspace{2mm}
\subsubsection{Analysis of $\mathcal S_{\gamma_2}^{\mathrm{c}}$}
We recall the Berry-Esseen CLT.
\vspace{1mm}
\begin{theo}[Berry-Esseen CLT $\mbox{\cite[Thm.~2]{erokhin1958varepsilon}}$]\label{berryessen}
Given $n>0$ and $Z_i$, $i=1, \ldots,n$  independent r.v.s. Then, for any real $t$,
\vspace{0.5mm}
 \begin{equation*}
  \left \lvert \mathbb{P} \left\{ \sum_{i=1}^n Z_i >n \left(\mu_n +t \sqrt{\frac{V_n}{n}}\right)\right\} -Q(t) \right \rvert \leq \frac{B_n}{\sqrt{n}},
 \end{equation*}where $\mu_n = \frac{1}{n} \sum_{i=1}^n \mathbb E [Z_i]$, $V_n  = \frac{1}{n} \sum_{i=1}^n \text{Var} [Z_i] $, $T_n =\frac{1}{n}\sum_{i=1}^n \mathbb{E} [{\lvert Z_i - \mu_i \rvert}^3] $, and $ B_n = 6 \frac{T_n}{V_n^{3/2}}$, and $Q(\cdot)$  is the tail distribution function of the standard normal distribution.
\end{theo}

We want to use Theorem \ref{berryessen} to estimate $\bar P_{W^n Y^n} {\left(\mathcal S_{\gamma_2}^{\mathrm{c}}  \right)}$, where $\mathcal S_{\gamma_2}$ is defined in \eqref{sgamma2}. We observe that, given $W^n=\mathbf w$, the terms
$Z_i= h_{\bar P}(w_i|Y_i)$ for $i=1, \ldots n$ are mutually independent because of the choice for the mismatch SLC \eqref{decoder}. 
 Then, if 
 \vspace{-2mm}
 {\allowdisplaybreaks
 \begin{align}
 \MoveEqLeft[5]
 n(R+R_0)>\underbrace{\sum_{i=1}^n { \mathbb E}_{\bar P_{Y_i|w_i}} [h_{\bar P}(w_i|Y_i)]}_{n \mu_n } \label{be rate}\\
&  + Q^{-1} (\varepsilon_4) \!\underbrace{\sqrt{ \!\sum_{i=1}^n \! {\text{Var}}_{\bar P_{Y_i|w_i}} (h_{\bar P}(w_i|Y_i))}}_{n \, \sqrt{ V_n/n}} +\, \gamma_2,\nonumber
 \end{align}\vspace{-2mm}}the chain of inequalities
  {\allowdisplaybreaks
 \begin{align*}
  \sum_{i=1}^n h_{\bar P}(W|Y) >n(R+R_0)-\gamma_2> n\left(\mu_n + t\sqrt{\frac{V_n}{n}}\right)
  \end{align*}}implies that $\mathcal S_{\gamma_2}^{\mathrm{c}}$ is contained in 
   {\allowdisplaybreaks
 \begin{align*}
 \left\{  (\mathbf w,\mathbf y) : \sum_{i=1}^n \! h_{\bar P}(w_i|y_i)> n\mu_n + n \, Q^{-1}(\varepsilon_4) \sqrt{\frac{V_n}{n}}\right\}\!.\stepcounter{equation}\tag{\theequation}\label{be2}
 \end{align*}}Therefore, if we apply Theorem \ref{berryessen} to the right-hand side of \eqref{be2}, and we choose 
 \begin{equation}\label{epsilon4}
 Q(t)=\varepsilon_4=\varepsilon_5 + \frac{B_n}{\sqrt{n}}, \vspace{-2mm}
 \end{equation}we have
    {\allowdisplaybreaks
 \begin{align}
 &\left\lvert  \mathbb P \left\{ \sum_{i=1}^n \! h_{\bar P}(w_i|y_i)>n\mu_n \!+\! n\, Q^{-1}(\varepsilon_4) \sqrt{\frac{V_n}{n}} \right\}\! -\! \varepsilon_4  \right\rvert \leq \! \frac{B_n}{\sqrt{n}},\nonumber\\[1mm]
&\,\,\, \mathbb P \left\{ \sum_{i=1}^n \! h_{\bar P}(w_i|y_i)>n\mu_n 
+ n\, Q^{-1}(\varepsilon_4) \sqrt{\frac{V_n}{n}}\right\} \leq \varepsilon_5.\label{be4}
 \end{align}}Finally, \eqref{be4} combined with \eqref{be2} implies $\bar P_{W^n Y^n}\!{\left(\mathcal S_{\gamma_2} ^{\mathrm{c}}  \right)} \leq \varepsilon_5.$
 
\vspace{2.2mm} 
\begin{rem}[Channel dispersion]
Observe that
\vspace{-2mm}
{\allowdisplaybreaks
 \begin{subequations}
 \begin{align}
 \mu_n &\coloneqq\frac{1}{n} \sum_{i=1}^n { \mathbb E}_{\bar P_{Y_i|w_i}} [h_{\bar P}(w_i|Y_i)]\nonumber\\[0.5mm]
& =  {\mathbb E}_{w} { \mathbb E}_{\bar P_{Y|W}} [h_{\bar P}(W|Y)|W]\nonumber\\[1mm]
& = {\mathbb E}_{(w,y)}  [h_{\bar P}(W|Y)|W]=H_{\bar P}(W|Y),\label{rate entropy}\\[2mm]
 V_n&\coloneqq  \frac{1}{n} \sum_{i=1}^n {\text{Var}}_{\bar P_{Y_i|w_i}} (h_{\bar P}(w_i|Y_i))\nonumber\\
& = \mathbb E_{w} \left[  {\text{Var}}_{\bar P_{Y|W}} (h_{\bar P}(W|Y)|W) \right]\nonumber\\
&= \mathbb E_{w} \left[  {\text{Var}}_{\bar P_{Y|W}} (\imath_{\bar P}(W;Y)|W) \right],\label{dispertion term}
\end{align} \end{subequations}}and $V_{\bar P}=\min_{\bar P_{W}} \mathbb E_{w} \left[  {\text{Var}}_{\bar P_{Y|W}} (\imath_{\bar P}(W;Y)|W) \right] $ is the dispersion of the channel $\bar P_{Y|W}$ as defined in \cite[Thm.~49]{polyanskiy2010channel}.
Hence, \eqref{be rate} can be rewritten as
 \begin{equation}\label{be rate3}
 n(R+R_0)>n H_{\bar P}(W|Y)  + n\, Q^{-1}(\varepsilon_4) \sqrt{\frac{V_{\bar P}}{n}} + \gamma_2.
 \end{equation}
\end{rem}


\subsubsection{Choice of $(\gamma_1, \gamma_2, \gamma_3)$} 

If we choose $(\gamma_1, \gamma_2, \gamma_3)=( \log{n},  \frac{1}{2} \log{n},  \log{n})$, \eqref{rateepsilon2}, \eqref{epsilon3}, and \eqref{be rate3} become
\vspace{-1mm}
{\allowdisplaybreaks
\begin{align}
 & R+R_0 > H_{\bar P}(W|Y) + Q^{-1}(\varepsilon_4) \sqrt{ \frac{V_{\bar P}}{n}}+ (\log{n}/2 \,n),\nonumber\\
 & R+R_0 < H_{\bar P}(W|U) - \varepsilon_2- (\log{n}/n) ,\nonumber\\
 & R< H_{\bar P}(W|UXYV) - \varepsilon_3- ( \log{n}/n),\label{rate2}
\end{align}}and the bound \eqref{epsilon tot}  on the $L^1$ distance becomes
\vspace{0.5mm}
{\allowdisplaybreaks
\begin{align}
 &{\lVert  P^{\text{RB}}_{U^n X^n Y^n V^n }- P^{\text{RC}}_{U^n X^n Y^n V^n} \rVert}_{1} \leq \varepsilon_{\text{Tot}} ,\nonumber\\[-0.5mm]
 & \varepsilon_{\text{Tot}} \!=\! 2 \bar P_{U^n CF}{ \left( \mathcal S_{\gamma_1}^{\mathrm{c}}  \right)}\!\! +\!  10 \bar P_{W^n Y^n}\!{\left(\mathcal S_{\gamma_2}^{\mathrm{c}}  \right)}\!\!+\!  2  P^{\text{RB}}\!{ \left( \mathcal S_{\gamma_3}^{\mathrm{c}}  \right)}\!\!+\! \frac{10 \!+\!\!2\sqrt{2}}{\sqrt{n}}\nonumber\\[-1mm]
& \phantom{ \varepsilon_{\text{Tot}} =} \leq  10\, \varepsilon_5 +\frac{10+2\sqrt{2}}{\sqrt{n}}.\label{speed conv}
\end{align}}

\vspace{-0.4cm}

\subsection{Rate conditions}\label{section rate}
With this choice for $\gamma_i$, \eqref{rate2} can be rewritten as:
\vspace{-1mm}
{\allowdisplaybreaks
\begin{align}
 & I_{\bar P}(W;U) < I_{\bar P}(W;Y)+ \varepsilon_2+\underbrace{\frac{3 \log{n}}{2\, n}}_{(\gamma_1+\gamma_2)/n}\! + Q^{-1}(\varepsilon_4) \sqrt{ \frac{V_{\bar P}}{n}},\nonumber\\[0.5mm]
 & R_0\!>I_{\bar P}(W;UXV|Y)\!+\! \varepsilon_3+\!\!\!\underbrace{\frac{3 \log{n}}{2\, n}}_{(\gamma_2\!+ \gamma_3)/n}\!\!+ Q^{-1}(\varepsilon_4) \sqrt{ \frac{V_{\bar P}}{n}},\label{final rate}
\end{align}}where $(\varepsilon_2,\varepsilon_3,\varepsilon_4)$ are defined in \eqref{epsilon2}, \eqref{epsilon3}, and \eqref{epsilon4}.

\vspace{2mm}
 \begin{rem}[Trade-off between  $\varepsilon_{\text{Tot}}$ and rate]
Observe that in order to minimize  $\varepsilon_{\text{Tot}}$, we can choose $\varepsilon_5$ equal to zero in \eqref{epsilon4} and \eqref{be4}. On the other hand,  this would require more common randomness since $Q^{-1}(\cdot)$ increases as its argument approaches zero.
Note that one can minimize $ \varepsilon_{\text{Tot}}$ (for example, we can have $ \varepsilon_{\text{Tot}}= \text{constant} \cdot e^{-n}$) simply by choosing greater $(\gamma_1, \gamma_2, \gamma_3)$ in  \eqref{epsilon tot},  but this increases the rate conditions~\eqref{final rate}.
\end{rem}

\begin{small}
\bibliographystyle{IEEEtran}
\bibliography{mybib}
\end{small}

\end{document}